\documentclass[twocolumn]{aastex63}
\usepackage{float,graphicx,amsmath,multirow}
\usepackage{color}
\usepackage[version=4]{mhchem}
\usepackage{booktabs}
\usepackage{gensymb}

\graphicspath{{figures/}}

\newcommand{\nth}{\textsuperscript{th}}

\begin{document}

\title{Discovery of Interstellar \emph{trans}-cyanovinylacetylene (HCCCH=CCHCN) and vinylcyanoacetylene (H$_2$C=CHC$_3$N) in GOTHAM Observations of TMC-1}
\author{Kin Long Kelvin Lee}
\affiliation{Department of Chemistry, Massachusetts Institute of Technology, Cambridge, MA 02139, USA}
\author{Ryan A. Loomis}
\affiliation{National Radio Astronomy Observatory, Charlottesville, VA 22903, USA}
\author{Andrew M. Burkhardt}
\affiliation{Center for Astrophysics $\mid$ Harvard~\&~Smithsonian, Cambridge, MA 02138, USA}
\author{Ilsa R. Cooke}
\affiliation{Univ Rennes, CNRS, IPR (Institut de Physique de Rennes)}
\author{Ci Xue}
\affiliation{Department of Chemistry, University of Virginia, Charlottesville, VA 22904, USA}
\author{Mark A. Siebert}
\affiliation{Department of Astronomy, University of Virginia, Charlottesville, VA 22904, USA}
\author{Christopher N. Shingledecker}
\affiliation{Benedictine College, Atchison, KS 66002, USA}
\author{Anthony Remijan}
\affiliation{National Radio Astronomy Observatory, Charlottesville, VA 22903, USA}
\author{Steven B. Charnley}
\affiliation{Astrochemistry Laboratory and the Goddard Center for Astrobiology, NASA Goddard Space Flight Center, Greenbelt, MD 20771, USA}
\author{Michael C. McCarthy}
\affiliation{Center for Astrophysics $\mid$ Harvard~\&~Smithsonian, Cambridge, MA 02138, USA}
\author{Brett A. McGuire}
\affiliation{Department of Chemistry, Massachusetts Institute of Technology, Cambridge, MA 02139, USA}
\affiliation{National Radio Astronomy Observatory, Charlottesville, VA 22903, USA}
\affiliation{Center for Astrophysics $\mid$ Harvard~\&~Smithsonian, Cambridge, MA 02138, USA}

\correspondingauthor{Kin Long Kelvin Lee, Brett A. McGuire}
\email{kelvlee@mit.edu, brettmc@mit.edu}

\begin{abstract}

We report the discovery of two unsaturated organic species, \textit{trans}-(E)-cyanovinylacetylene and vinylcyanoacetylene, using the second data release of the GOTHAM deep survey towards TMC-1 with the 100\,m Green Bank Telescope. For both detections, we performed velocity stacking and matched filter analyses using Markov chain Monte Carlo simulations, and for \textit{trans}-(E)-cyanovinylacetylene,  three rotational lines were observed at low signal-to-noise (${\sim}$3$\sigma$). From this analysis, we derive column densities of $2\times10^{11}$ and $3\times10^{11}$\,cm$^{-2}$ for vinylcyanoacetylene and \textit{trans}-(E)-cyanovinylacetylene, respectively, and an upper limit of $<2\times10^{11}$\,cm$^{-2}$ for \textit{trans}-(Z)-cyanovinylacetylene. Comparisons with G3//B3LYP semi-empirical thermochemical calculations indicate abundances of the [\ce{H3C5N}] isomers are not consistent with their thermodynamic stability, and instead their abundances are mainly driven by dynamics. We provide discussion into how these species may be formed in TMC-1, with reference to related species like vinyl cyanide (\ce{CH2=CHC#N}). As part of this discussion, we performed the same analysis for ethyl cyanide (\ce{CH3CH2C#N}), the hydrogenation product of \ce{CH2=CHC#N}. This analysis provides evidence---at 4.17$\sigma$ significance---an upper limit to the column density of $<4\times10^{11}$\,cm$^{-2}$; an order of magnitude lower than previous upper limits towards this source.

\end{abstract}

\section{Introduction}
\label{intro}

Radio observations of the Taurus Molecular Cloud (TMC) complex, particularly towards the prestellar cloud core TMC-1, have revealed a plethora of molecules ranging from the cyanopolyynes (\ce{HC_nN}, for odd values of $n$), to free radicals [e.g., \textit{l}-\ce{C3H} by \cite{thaddeus_astronomical_1985}], to carbenes (e.g. \ce{H2C6} by \cite{langer_first_1997}), to ions (e.g., \ce{C8H-} by \cite{brunken_detection_2007}). More recently, the discovery of benzonitrile (\textit{c}-\ce{C6H5CN}) by \citet{mcguire_detection_2018} adds an aromatic ring--- noteworthy for its exceptional thermodynamic and chemical stability, and a key building block in biological systems and the formation of soot and interstellar dust---to this already rich inventory. Chemical models, however, currently do not have a sufficiently efficient pathway to reproduce the abundance of these aromatic molecules, in part due to the lack of observational constraints on potential carbon-chain precursors \citep{burkhardt_ubiquitous_2020}. Unlike the well-studied cyanopolyyne family, many of the partially saturated carbon-chains have never been detected and thus large unknowns for these models. As such, in order to determine the formation of even the simplest aromatics, robust abundance measurements of partially saturated must be obtained. Understanding how molecules like benzonitrile may be formed, processed, and transported in the interstellar medium---in particular in cold, dark environments like TMC-1---has significant implications in the chemical evolution of these environments.

Our large-scale observing campaign with the 100\,m Green Bank Telescope, GOTHAM (GBT Observations of TMC-1: Hunting for Aromatic Molecules), seeks to determine the chemical inventory of TMC-1 at an unprecedented level by performing a wide band spectral line survey at centimeter wavelengths at high uniform sensitivity (target 2\,mK RMS across the whole spectrum) and high resolution (0.05\,km s$^{-1}$). As part of the first data release,  molecules of considerable complexity were reported for the first time, including 1- and 2-cyanonaphthalene (\ce{C10H7CN}) \citep{mcguire_discovery_2020}, \textit{c}-\ce{C5H5CN} \citep{mccarthy_interstellar_2020}, \ce{HC11N} \citep{loomis_investigation_2020}, and \ce{HC4NC} \citep{xue_detection_2020}.  Many of these detections greatly benefited from  combining signal processing techniques with Bayesian modeling,  which enables identification of molecules in sparse line spectra even when no individual features are present above the noise.  Furthermore, statistically robust derivations of parameters such as column density and excitation temperature can be determined using this treatment. For an overview of this method and its use cases, the reader is referred to \citet{mcguire_early_2020}.

In this paper, we examine evidence in the second data release of GOTHAM for three isomers in the \ce{H3C5N} family: vinylcyanoacetylene (\ce{H2C=CHC3N}, VCA), and the ($E$) and ($Z$) isomers of \textit{trans}-cyanovinylacetylene (\ce{HC#CCH=CHC#N}, CVA), as shown in Fig.~\ref{fig:energetics}. They are extended variants of vinyl cyanide (\ce{CH2CHCN}), an unsaturated nitrile-bearing molecule which was first detected in this source by \cite{matthews_detection_1983}. Given questions still persist as to how small branched hydrocarbon chains form in TMC-1 despite their structural  simplicity \citep{vigren_dissociative_2009}, simultaneous analysis of multiple isomers may provide insight into the operative formation pathways there. 
Of particular interest for this isomeric family is the possible connection with aromatic N-heterocycles such as pyridine, $c$-\ce{C5H5N}, which might be formed by subsequent hydrogenation of one or more of the these isomers followed by ring closure.

\section{Observations}

Observations with the 100\,m Green Bank Telescope (GBT) were carried out for the GOTHAM project, which has been detailed in a series of publications. Briefly, this work uses the second data release (henceforth referred to as DR2) of GOTHAM, which are observations targeting the well-known TMC-1 cyanopolyyne peak (CP) centered at $\alpha_\mathrm{J2000}$ = 04\fh41\fm42.5\fs, $\delta$ = +25\arcdeg41\arcmin26.8\arcsec \citep{mcguire_early_2020}. As of DR2, our spectra cover the GBT X-, K-, and Ka-bands from 7.906 to 33.527 GHz (25.6 GHz bandwidth) with continuous coverage between 22--33.5 GHz, at a uniform frequency resolution of 1.4\,kHz (0.05--0.01\,km/s in velocity) and an RMS noise of ${\sim}2$--20\,mK across the spectrum, with the RMS increasing towards higher frequency due to limited integration time.  Uncertainty due to flux calibration is expected to be $\sim$20\%, based on complementary VLA observations of the flux-calibrator source J0530+1331 \citep{mcguire_early_2020}.

\section{Computational methods}

\subsection{Quantum chemistry \label{sub:qchem}}

Calculations of the permanent electric dipole moment and relative energetics were performed using the Gaussian \textsc{\char13}16 suite of electronic structure programs \citep{frisch_gaussian_2016}. For dipole moments, geometries were first optimized at the $\omega$B97X-D/6-31+G(d) level of theory, and the one-electron properties calculated at the same level---based on our earlier benchmarking, this combination produces dipole moments with uncertainties on the order of $\pm0.5$\,D and systematically over-predicts vibrationally averaged values by 0.1\,D \citep{lee_bayesian_2020}. In terms of thermochemistry, we used the B3LYP variant of the G3 semi-empirical model chemistry \citep{baboul_gaussian-3_1999}, which has been shown to be a computationally cost-effective method of obtaining near-chemically accurate energetics (${\sim}$120\,K)  \citep{simmie_benchmarking_2015}.

\subsection{MCMC modeling}

Details of the Markov chain Monte Carlo (MCMC) simulations are described in depth by \citet{loomis_investigation_2020}, and we only briefly discuss the relevant aspects here. The objective of these simulations is to properly model the physical parameters of molecules in TMC-1 where maximum likelihood methods generally fail due to high covariance and dimensionality. The model used in our study assumes a Gaussian shape for the spatial distribution of TMC-1, with parameters for the size of the source ($SS$), radial velocity ($v_\mathrm{LSRK}$), column density ($N_\mathrm{col}$), excitation temperature ($T_\mathrm{ex}$), and spectral linewidth ($dv$). Based on recent observations performed with the 45\,m telescope at Nobeyama Radio Observatory \citep{dobashi_spectral_2018,dobashi_discovery_2019,soma_complex_2018}, as well as our data \citep{loomis_investigation_2020}, emission from molecules in TMC-1 towards the cyanopolyyne peak display at least four individual velocity components which are subsequently represented by independent source size, radial velocity, and column density parameters; in total, our model comprises 14 fitting parameters.

Line profile simulations were performed using \mbox{\textsc{molsim}} \citep{lee_molsim_2020}. The MCMC simulations used wrapper functions in \textsc{molsim} to \textsc{arviz} \citep{kumar_arviz_2019} and \textsc{emcee} \citep{foreman-mackey_emcee_2013}; the former for analyzing the results of sampling, and the latter implements an affine-invariant MCMC sampler. As a prior parameters, we used the marginalized posterior from modeling \ce{HC9N} chosen based on chemical similarity. The prior distributions approximated as normal (i.e. $p(\theta)\sim N(\mu_\theta, \sigma_\theta)$ for parameter $\theta$) with modifications to the variance $\sigma_\theta$ as to avoid overly constrictive/influential priors. The MCMC analysis was performed first for \textit{trans}-(E)-CVA, which produced the strongest response out of the molecules reported here. Convergence of the MCMC was confirmed using standard diagnostics such as the \citet{gelman_single_1992} $\hat{R}$ statistic, and by visually inspecting the posterior traces. The resulting posterior for \textit{trans}-(E)-CVA was subsequently used as prior distributions for VCA and \textit{trans}-(Z)-CVA, albeit with the source sizes constrained to the mean values of \textit{trans}-(E)-CVA due to poorly convergent sampling. The results reported here are relatively insensitive to the choice of prior; comparison between benzonitrile and \textit{trans}-(E)-CVA do not qualitatively change the statistics derived from the converged posteriors.

\subsection{Velocity stacking and matched filter analysis}

With the model posterior at hand, we can corroborate our simulations with the observed data through a combined velocity stack and matched filter analysis \citep{loomis_detecting_2018,loomis_investigation_2020}. Briefly, the former involves a noise-weighted composite spectrum by stacking the \emph{observed} spectra---in velocity space---using known molecular transition frequencies, and the latter performs a velocity stack of the spectral simulation based on the fitted parameters, and cross-correlated with the observational velocity stack. The advantage of this approach is the ability to derive statistical significance of a detection based on the impulse response in the cross-correlation: the significance, $\sigma$, directly quantifies how well our model reproduces the observed data in velocity space, even in lieu of individual observed transitions. Throughout the GOTHAM studies, we have adopted a $\geq$5$\sigma$ heuristic for what constitutes a molecular detection as determined by \ce{HC11N} \citep{loomis_investigation_2020}.

\section{Results \& discussion}

\subsection{Spectroscopy \& relative energetics}

\begin{figure}
    \centering
    \includegraphics[width=0.49\textwidth]{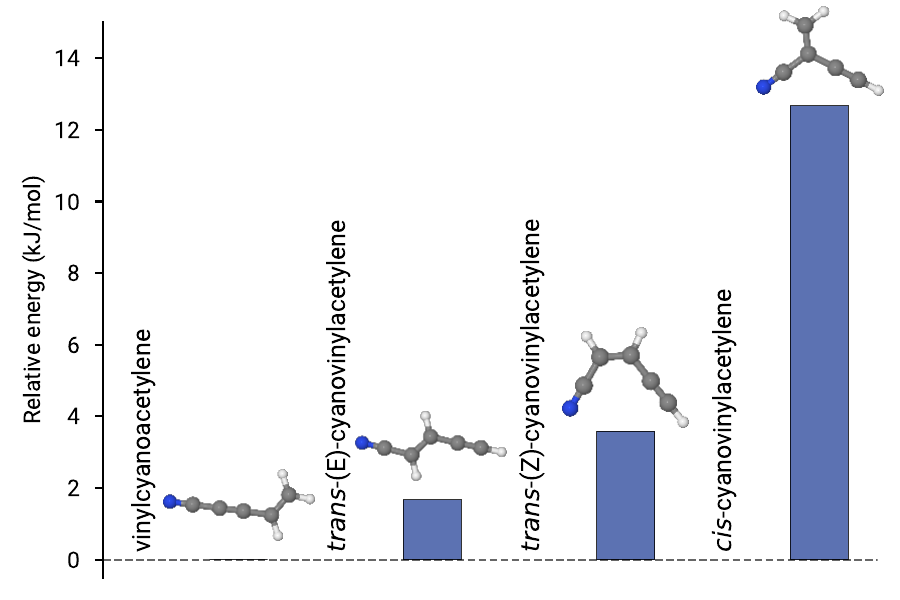}
    \caption{G3//B3LYP energetics of the [\ce{H3C5N}] structures of interest at 0\,K, given relative to the lowest energy form vinylcyanoacetylene.}
    \label{fig:energetics}
\end{figure}

Figure \ref{fig:energetics} shows the relative energetics of the four molecules under investigation, along with their equilibrium structures. The two lowest energy forms of \ce{[H3C5N]} are near-prolate tops, while the two higher energy forms are closer to the oblate limit. The highest energy form in our study, \textit{cis}-cyanovinylacetylene, has not yet been experimentally observed whereas the remaining three isomers have been studied extensively in the laboratory \citep{halter_microwave_2001,thorwirth_high_2004}, and most recently observed in a discharge mixture of benzene and \ce{N2} \citep{mccarthy_exhaustive_2020}. We have also omitted the linear chain form, \ce{CH3C4N}, which is likely to be unstable by analogy to the cyanpolyynes.

Catalogs of rotational transitions were generated using the SPCAT program \citep{pickett_fitting_1991} based on spectroscopic parameters---including nitrogen-hyperfine splitting---reported in the cited publications. In all cases, our electronic structure calculations suggest sizable dipole moments along $a$ and $b$-inertial axes (Table \ref{tab:dipoles}) with the total dipole moment around ${\sim}5$\,D; typical for \ce{-C#N} bearing molecules. For the two molecules closer to the oblate limit, the total dipole moment is divided roughly equally between the $a$ and $b$-axes. Given that $I\propto \mu^2$, this means that on an individual line basis, the near-oblate isomers require an order of 2--3 times more sensitivity for detection compared to their near-prolate counterparts.

\begin{deluxetable}{lrr}
\tablecaption{Theoretical equilibrium dipole moments for the molecules of interest. Values are calculated at the $\omega$B97X-D/6-31+G(d) level of theory, which have a nominal $\pm$0.5\,D uncertainty (see Section \ref{sub:qchem}). \label{tab:dipoles}}
\tablehead{
    ~ & \colhead{$\mu_a$} & \colhead{$\mu_b$} \\
    ~ & \colhead{(Debye)} & \colhead{(Debye)} 
    }
\startdata
VCA & 5.3 & 0.3 \\
\textit{trans}-(E)-CVA & 4.2 & 0.6 \\
\textit{trans}-(Z)-CVA & 2.7 & 2.8 \\
\textit{cis}-CVA & 3.2 & 2.5 \\
\enddata
\end{deluxetable}

\subsection{GOTHAM observations}

\begin{figure}
    \centering
    \includegraphics[width=\columnwidth]{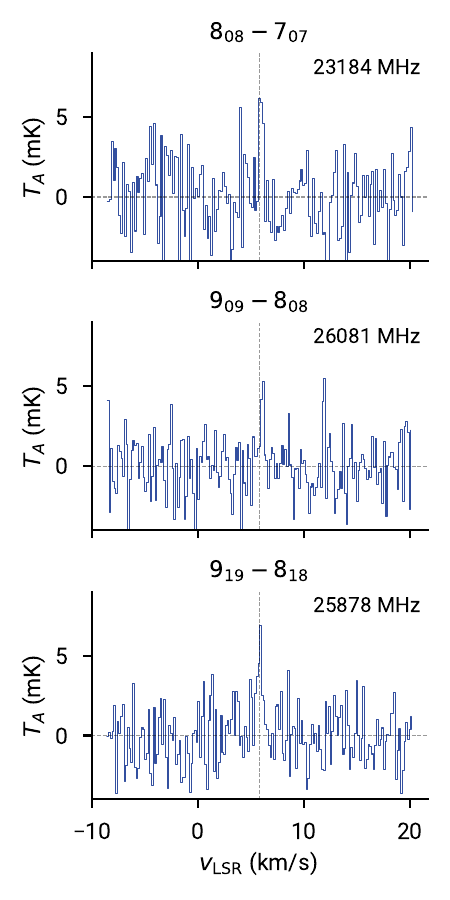}
    \caption{Observed spectral features in the GOTHAM spectrum (rebinned to 14 kHz resolution) at the center frequencies for three \textit{a}-type transitions of \textit{trans}-(E)-CVA. The dashed vertical line indicates the nominal source velocity at 5.8 km/s. Asymmetric top ($J_{K_a K_c}$) quantum number assignments are given at the top of each trace, along with their rest frequencies. The peak signal-to-noise ratio for each spectrum is on the order of ${\sim}$3$\sigma$.}
    \label{fig:lines}
\end{figure}

Based on the energetics shown in Figure \ref{fig:energetics}, we attempted to search for individual transitions arising from the lower energy isomers. Upon inspection, the lowest energy isomer, VCA, does not exhibit any obvious features in our spectra. For the next isomer in energy, \textit{trans}-$(E)$-CVA, Figure \ref{fig:lines} shows three spectral windows centered at frequencies that correspond to two $K=0$ and one $K=1$ transitions. These windows hint at individual spectral features within our data, albeit at low significance.

\subsection{Velocity stack and matched filter analysis}

\begin{figure}
    \centering
    \includegraphics[width=\columnwidth]{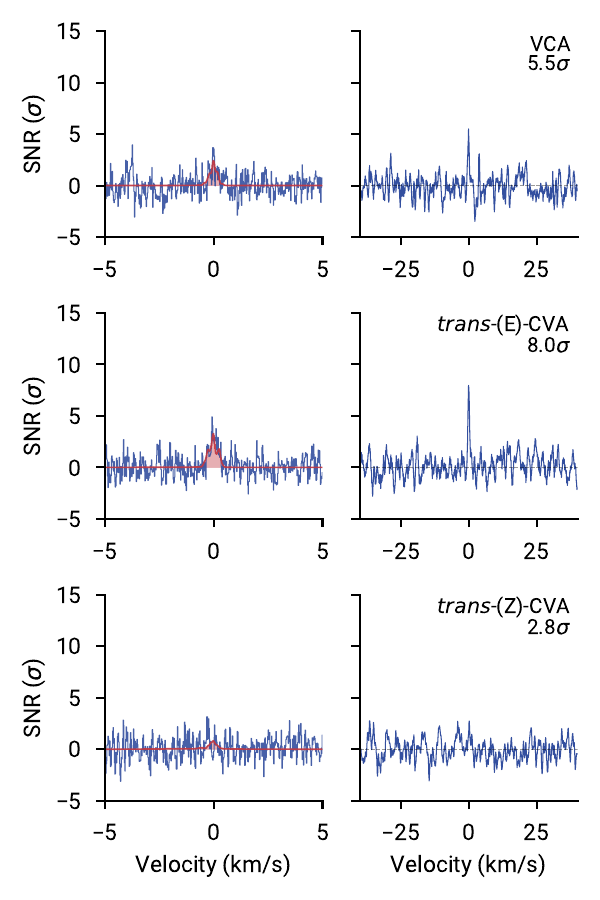}
    \caption{Velocity stack (left columns) and corresponding matched filter (right columns) for the three molecules of interest. Peak impulse responses---shown in units of signal-to-noise (SNR)---are provided in the matched filter plots. The velocity stack of simulated spectra using parameters derived from the MCMC simulations are overlaid in red.}
    \label{fig:mf-plots}
\end{figure}

From our observations, only \textit{trans}-(E)-CVA exhibits tentative individual spectral features---whereas conventional methods of analysis (i.e. least-squares fits) may be limited to deriving upper limits to column densities, here we can combine velocity stacking and matched filter analysis with MCMC simulations to derive statistically robust molecular parameters for all three isomers. Figure \ref{fig:mf-plots} visualizes the detection of VCA and \textit{trans}-(E)-CVA, and non-detection of \textit{trans}-(Z)-CVA with the velocity stack and matched filter analyses. 

The velocity stacks in Figure \ref{fig:mf-plots} indicate the presence of additive spectral intensity across the GOTHAM survey for VCA and \textit{trans}-(E)-CVA, and none for \textit{trans}-(Z)-CVA. For each stack, the red traces show the velocity stack for simulated spectra based on the posterior means from the MCMC simulations, which in all cases agrees well with the stacks based on observations. The two velocity stacks corroborate in the matched filter, simultaneously visualizing the overlap between model and observations, and providing an estimate of the significance of our detection. For VCA and \textit{trans}-(E)-CVA, the matched filters exhibit a response clearly above the noise, whereas for \textit{trans}-(Z)-CVA this is not the case and thus we report only an upper limit to the column density for this molecule. Table \ref{tab:columns} summarizes the derived column densities and excitation temperatures from the MCMC simulations.

\begin{deluxetable}{lcc}
\tablecaption{Total column densities and excitation temperatures derived from the MCMC simulations. Uncertainties are given as the 95\% credible interval. Unabridged MCMC results can be found in the Appendix. Interlopers refer to interfering features of other species, detected as $3\sigma$ for a given spectral chunk. \label{tab:columns}}
\tablehead{
    \colhead{Molecule} & \colhead{Column density} & \colhead{$T_\mathrm{ex}$} \\
    ~ & \colhead{(10$^{11}$ cm$^{-2}$)} & \colhead{(K)}
    }
\startdata
VCA\tablenotemark{a} & 1.87$^{+0.37}_{-0.38}$ & 6.7$^{+0.6}_{-0.6}$ \\
\textit{trans}-(E)-CVA\tablenotemark{b} & 2.90$^{+0.41}_{-0.40}$ & 7.0$^{+1.2}_{-1.3}$\\
\textit{trans}-(Z)-CVA\tablenotemark{c} & $<2$\tablenotemark{d} & --- \\
\enddata
\tablenotetext{a}{Based on 271 transitions, with zero ignored due to interlopers.}
\tablenotetext{b}{Based on 270 transitions, with one ignored due to interlopers.}
\tablenotetext{c}{Based on 1354 transitions, with zero ignored due to interlopers.}
\tablenotetext{d}{Upper limit given as the 97.5\nth~percentile.}
\end{deluxetable}

\subsection{Astrochemical implications}

Based on our systematic investigation of the three lowest energy isomers of \ce{[H3C5N]}, we can infer some details into the relative importance of dynamical pathways that lead to the overall chemical inventory of TMC-1---particularly the branched unsaturated hydrocarbons we see here. In all cases, the excitation temperatures we estimate are on the order of ${\sim}7$\,K---consistent with those reported for other molecules in TMC-1 \citep{dobashi_spectral_2018,dobashi_discovery_2019}. The velocity profiles are also similar between each isomer, allowing us to assume that they are cospatial within each velocity component. Of the two isomers we successfully detected, VCA is approximately two times less abundant than \textit{trans}-(E)-CVA, despite being more thermodynamically stable (${\sim}$240\,K), thus their relative abundance is dominated by dynamics. The third isomer, \textit{trans}-(Z)-CVA, does not exhibit a significant response in either the velocity stack nor the matched filter (Figure \ref{fig:mf-plots}). However, our MCMC simulations indicate that it may be just below our current sensitivity limits (see Appendix \ref{fig:transZCVA-corner}).

Given that the relative abundances are likely dictated by kinetics, the question now turns to how the three isomers may be preferentially formed in TMC-1. In the gas-phase, neutral-neutral reactions are an attractive route: radicals such as \ce{C2H} and \ce{CN} can react with \ce{CH2=CHC#N} (vinyl cyanide) and \ce{CH2=CHC#CH} (vinyl acetylene), followed by hydrogen loss. The reaction \ce{C2H + CH2=CHC#N} and would produce stereoisomers of CVA, with a preference for the \textit{trans} isomers due to steric hinderance owing to the acetylenic unit. While this specific reaction has not yet been studied experimentally, by analogy to similar reactions between \ce{C2H} and unsaturated hydrocarbons [\ce{C2H2} \citep{kovacs_h-atom_2010,chastaing_neutralneutral_1998,zhang_crossed_2009}]; \ce{C2H4}, \ce{C3H6} \citep{bouwman_bimolecular_2012,krishtal_theoretical_2009,chastaing_neutralneutral_1998}] we expect this reaction to be barrierless and efficient even at low temperatures. The latter reaction, \ce{CN + CH2=CHC#CH}, can form all three [\ce{H3C5N}] isomers considered in Figure \ref{fig:energetics}; experimental studies by \citet{yang_cn_1992,sims_rate_1993} suggests \ce{CN} addition is just as efficient to either the vinyl (forming CVA) or acetylenic unit (forming VCA) \citep{balucani_formation_2000,choi_h_2004} . Additionally, VCA can be formed through the reaction between \ce{C3N + C2H4} involving submerged barriers \citep{moon_reaction_2017}; experimental kinetic data for this reaction is not yet available to the best of our knowledge.

Alternatively, grain surface reactions are a well-established pathway to hydrogenate unsaturated species efficiently \citep{herbst_chemistry_2001,cuppen_grain_2017}. In this context, \ce{[H3C5N]} molecules would be formed through \ce{HC5N + H2} hydrogenation, with the isomeric ratio dependent on the relative cross-section or likelihood of attaching \ce{H2} to each respective part of the chain, which in turn is dictated by whether this occurs in a concerted (\ce{+H2}) or stepwise (\ce{+ H + H}) fashion. The former is likely to be highly endothermic and therefore unlikely under cold, dark conditions, while the latter is facilitated by hydrogen atom tunneling. A similar route had been proposed for \ce{CH2CHCN}, with \ce{HC3N} as the precursor \citep{blake_molecular_1987}. Analysis by \citet{loomis_investigation_2020} suggests \ce{HC5N} is approximately an order of magnitude less abundant than \ce{HC3N}, which corresponds well with isomers of the \ce{[H3C5N]} family and \ce{CH2CHCN}, where the column density of the latter was determined by \citet{matthews_detection_1983} to be on the order of 3$\times10^{12}$\,cm$^{-2}$. An argument against hydrogenation reactions, however, is that prior to sublimation from the grain they should ultimately form saturated species that are known to be uncommon in TMC-1: for instance, ethyl cyanide (\ce{CH3CH2CN}) should form from hydrogenation of \ce{CH2CHCN} \citep{blake_molecular_1987}. \citet{minh_upper_1991} were only able to place upper limits on \ce{CH3CH2CN}; using the same velocity stack and matched filter methodology, we tentatively ascribe an upper limit to the total column density of $<4\times10^{11}$\,cm$^{-2}$ (given as the 97.5th percentile; see Appendix \ref{sec:ethyl}), consistent with their determination ($<3\times10^{12}$\,cm$^{-2}$).

\begin{figure}[h]
    \centering
    \includegraphics[width=\columnwidth]{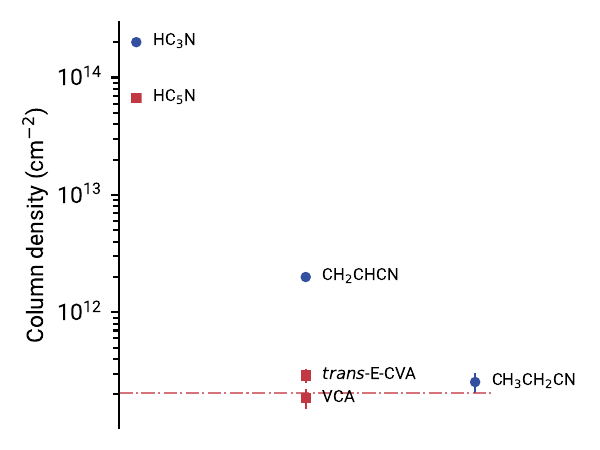}
    \caption{Column densities for assumed co-spatial \ce{HC3N} and \ce{HC5N} \citep{loomis_investigation_2020}, \ce{CH2CHCN} \citep{matthews_detection_1983}, \ce{CH3CH2CN} (Appendix \ref{sec:ethyl}), and the newly detected molecules toward TMC-1. The error bars represent 1$\sigma$ uncertainty.}
    \label{fig:decrement}
\end{figure}

While it is difficult to draw conclusions with confidence at this level of significance, it appears that the decrement in column density follows a qualitative trend (Figure \ref{fig:decrement}) that lends credit to sequential hydrogenation (\ce{HC3N ->[+H2] \ce{CH2CHCN} ->[+H2] \ce{CH3CH2CN}}). Better constraints on \ce{CH3CH2CN}, as well as the missing isomer \textit{trans}-(Z)-CVA will provide critical insight into the relative importance of gas phase and grain hydrogenation pathways.

If the hydrogenation route is indeed a large contributing mechanism, then VCA and \textit{trans}-(E)-CVA would be important intermediates toward the formation of cyclic molecules---specifically the still-elusive N-heterocycles such as pyrrole and pyridine, and the recently detected 1-cyanocyclopentadiene. In the former, we note that \textit{trans}-(Z)-CVA is a hydrogenation (\ce{+H2}) and ring closing step from pyridine---a molecule of biological importance. Further study into this isomeric family, particularly \textit{cis}-CVA and the deuterated isotopologues (similar to previous work on cyanopolyyne isotopologues by \cite{burkhardt_detection_2018}), should reveal the dynamical processes behind the formation of these molecules, although their laboratory spectra have not yet been measured.

\section{Conclusions}

From our GOTHAM observations, we report the first detection of two new isomers of [\ce{H3C5N}] toward TMC-1, and more generally in the interstellar medium. Combining MCMC simulations with velocity stacking and matched filter analysis, we were able to successfully characterize vinylcyanoacetylene (VCA) and \textit{trans}-(E)-cyanovinylacetylene (CVA) at 5.5$\sigma$ and 8.0$\sigma$ significance respectively, with derived column densities on the order of $1\times10^{11}$ and $2\times10^{11}$\,cm$^{-2}$ respectively. The third isomer, \textit{trans}-(Z)-CVA, appears to be just out of reach at current integration levels, and from our MCMC analysis we place an upper limit for its column density at $<2\times10^{11}$\,cm$^{-2}$. While it remains unclear how these unsaturated hydrocarbons may be formed in TMC-1, we discuss implications of cyanopolyyne hydrogenation on grain surfaces---as part of this analysis, we also report a tentative detection of ethyl cyanide with a MCMC derived upper limit to the total column density of $<4\times10^{11}$\,cm$^{-2}$. Further analysis into the \ce{[H3C5N]} family, and other related hydrocarbon chains will help reveal the complex formation processes taking place, and more broadly, how molecules more saturated than cyanopolyynes could be formed under cold, dark conditions.

\section{Data access \& code}

Data used for the MCMC analysis can be found in the DataVerse entry \citep{DVN/K9HRCK_2020}. The code used to perform the analysis is part of the \textsc{molsim} open-source package; an archival version of the code can be accessed at \cite{lee_molsim_2020}.

\acknowledgments

The National Radio Astronomy Observatory is a facility of the National Science Foundation operated under cooperative agreement by Associated Universities, Inc.  The Green Bank Observatory is a facility of the National Science Foundation operated under cooperative agreement by Associated Universities, Inc. M.C.M, and P.B.C. acknowledge financial support from NSF grants AST-1908576, AST-1615847, and NASA grant 80NSSC18K0396.  A.M.B. acknowledges support from the Smithsonian Institution as a Submillimeter Array (SMA) Fellow. I.R.C. acknowledges funding from the European Union’s Horizon 2020 research and innovation programme under the Marie Skłodowska-Curie grant agreement No 845165-MIRAGE. C.X. is a Grote Reber Fellow, and support for this work was provided by the NSF through the Grote Reber Fellowship Program administered by Associated Universities, Inc./National Radio Astronomy Observatory and the Virginia Space Grant Consortium.  

\bibliography{main}
\bibliographystyle{aasjournal}

\appendix

\renewcommand{\thefigure}{A\arabic{figure}}
\renewcommand{\thetable}{A\arabic{table}}
\renewcommand{\theequation}{A\arabic{equation}}
\setcounter{figure}{0}
\setcounter{table}{0}
\setcounter{equation}{0}

\section{MCMC posterior analysis}

\subsection{\textit{trans}-(E)-CVA}

\begin{figure}[h]
    \centering
    \includegraphics[width=\textwidth]{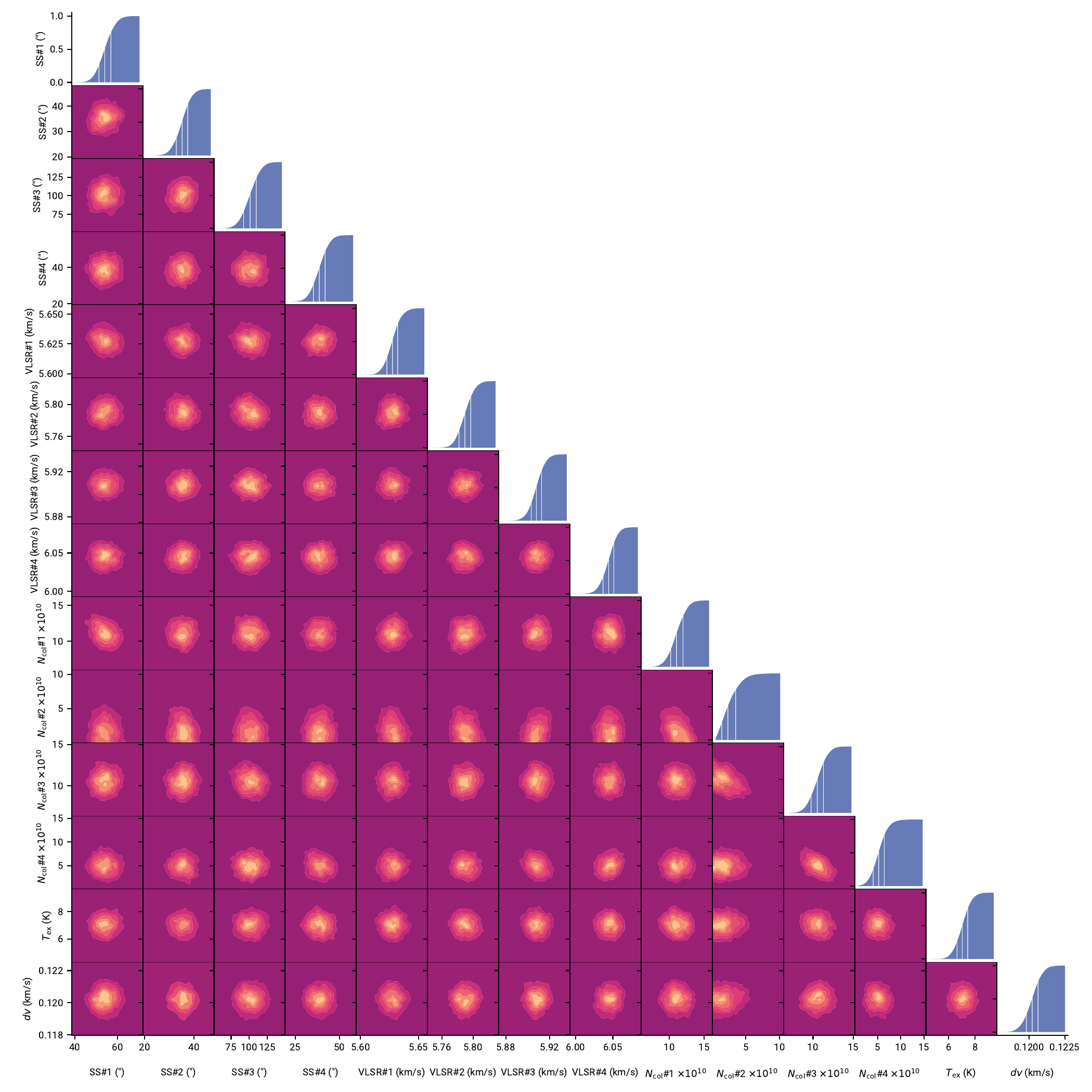}
    \caption{Corner plot for \textit{trans}-(E)-CVA. The diagonal traces correspond to ECDF plots, and off-diagonal plots show the kernel density covariance between model parameters. In the former, lines represent the 25{\nth}, 50{\nth}, and 75{\nth} percentiles respectively. The length scale for the kernel density plots is chosen with Scott's rule.}
    \label{fig:transECVA-corner}
\end{figure}

Figure \ref{fig:transECVA-corner} shows the results of the MCMC fit for \textit{trans}-(E)-CVA. There are several factors that warrant extra discussion, particularly as to how these plots can be interpreted. First, the diagonal traces correspond to the marginalized likelihood for each model parameter presented as empirical cumulative distribution function (ECDF) plots---these provide a non-parametric visualization of the likelihood, in contrast to kernel density estimates and histograms which require length scale and bin width specification respectively. Second, the off-diagonal traces correspond to kernel density plots of parameter pairs---these plots visualize the covariance between any given pair of model parameters; in our case,  well-approximated by two-dimensional Gaussian distributions.

Inspection of the marginalized likelihood ECDF traces indicate firm detections in components \#1, \#3, and \#4, with a likely non-detection in component \#. In the non-detection case, the cumulative density rises linearly from zero $N_\mathrm{col}$, while for detections they appear clearly sigmoid-like with the inflexion point at non-zero column. Table \ref{table:ecva} provides summary statistics for the posterior distributions.

\begin{table*}[hbt!]
\centering
\caption{\textit{trans}-(E)-CVA best-fit parameters from MCMC analysis. Quoted uncertainties correspond to the 95\% highest posterior density. $N_T$ (Total) refers to the pooled column density from the four components, given as the posterior mean and 95\% highest posterior density of the joint distributions.}
\label{table:ecva}
\begin{tabular}{c c c c c c}
\toprule
\multirow{2}{*}{Component}&	$v_{lsr}$					&	Size					&	\multicolumn{1}{c}{$N_\mathrm{col}$}					&	$T_{ex}$							&	$\Delta V$		\\
			&	(km s$^{-1}$)				&	($^{\prime\prime}$)		&	\multicolumn{1}{c}{(10$^{10}$ cm$^{-2}$)}		&	(K)								&	(km s$^{-1}$)	\\
\midrule
\hspace{0.1em}\vspace{-0.5em}\\
C1	&	$5.628^{+0.014}_{-0.015}$	&	$54^{+8}_{-9}$	&	$11.06^{+2.51}_{-2.59}$	&	\multirow{6}{*}{$7.0^{+1.2}_{-1.3}$}	&	\multirow{6}{*}{$0.120^{+0.001}_{-0.001}$}\\
\hspace{0.1em}\vspace{-0.5em}\\
C2	&	$5.790^{+0.022}_{-0.021}$	&	$35^{+7}_{-7}$	&	$2.31^{+2.78}_{-2.31}$	&	&	\\
\hspace{0.1em}\vspace{-0.5em}\\
C3	&	$5.908^{+0.015}_{-0.015}$	&	$101^{+26}_{-25}$	&	$10.52^{+2.27}_{-2.17}$	&	&	\\
\hspace{0.1em}\vspace{-0.5em}\\
C4	&	$6.044^{+0.021}_{-0.021}$	&	$39^{+10}_{-10}$	&	$5.04^{+3.72}_{-3.52}$	&	&	\\
\hspace{0.1em}\vspace{-0.5em}\\
\midrule
$N_T$ (Total)	&	 \multicolumn{5}{c}{$28.98^{+4.13}_{-3.98}\times 10^{10}$~cm$^{-2}$}\\
\bottomrule
\end{tabular}
\end{table*}

\newpage

\subsection{\textit{trans}-(Z)-CVA}

\begin{figure}[h]
    \centering
    \includegraphics[width=\textwidth]{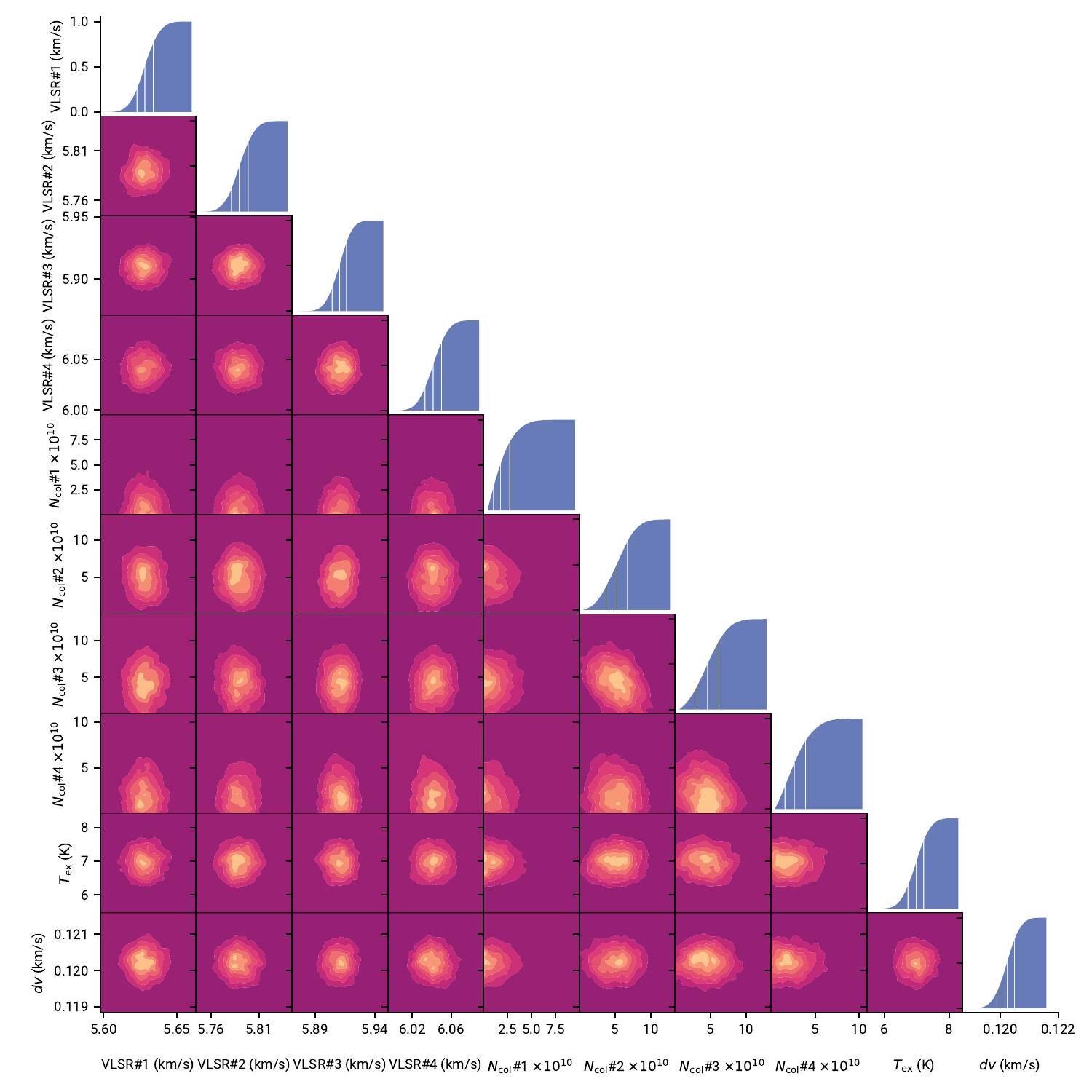}
    \caption{Corner plot for \textit{trans}-(Z)-CVA. The diagonal traces correspond to ECDF plots, and off-diagonal plots show the kernel density covariance between model parameters. In the former, lines represent the 25{\nth}, 50{\nth}, and 75{\nth} percentiles respectively. The length scale for the kernel density plots is chosen with Scott's rule.}
    \label{fig:transZCVA-corner}
\end{figure}

Figure \ref{fig:transZCVA-corner} shows the corner plot for \textit{trans}-(Z)-CVA. Because these simulations fix the source size to the mean of \textit{trans}-(E)-CVA, the source sizes are not sampled/fit and are omitted from the corner plot. Under these conditions, we observe most likely non-detection in component \#1, with evidence for detection in the remaining three components. Given that components \#2 and \#3 in particular are highly indicative of \textit{trans}-(Z)-CVA, we believe that this molecule is just out of reach at the current level of integration of the second GOTHAM data release. Summaries of the posterior distributions can be found in Table \ref{table:zcva}.

\begin{table*}[hbt!]
\centering
\caption{\textit{trans}-(Z)-CVA best-fit parameters from MCMC analysis. Quoted uncertainties correspond to the 95\% highest posterior density. $N_T$ (Total) refers to the pooled column density from the four components, given as the mean and 95\% highest posterior density of the joint posterior distributions.}
\label{table:zcva}
\begin{tabular}{c c c c c}
\toprule
\multirow{2}{*}{Component}&	$v_{lsr}$ &	\multicolumn{1}{c}{$N_\mathrm{col}$}					&	$T_{ex}$							&	$\Delta V$		\\
			&	(km s$^{-1}$)				&	\multicolumn{1}{c}{(10$^{10}$ cm$^{-2}$)} 	&	(K) & (km s$^{-1}$)	\\
\midrule
\hspace{0.1em}\vspace{-0.5em}\\
C1	&	$5.628^{+0.017}_{-0.016}$	&	$1.72^{+2.45}_{-1.72}$	&	\multirow{6}{*}{$7.0^{+0.7}_{-0.7}$}	&	\multirow{6}{*}{$0.120^{+0.001}_{-0.001}$}\\
\hspace{0.1em}\vspace{-0.5em}\\
C2	&	$5.789^{+0.025}_{-0.026}$	&	$5.07^{+4.15}_{-4.11}$	&	&	\\
\hspace{0.1em}\vspace{-0.5em}\\
C3	&	$5.910^{+0.017}_{-0.018}$	&	$4.47^{+3.77}_{-4.46}$	&	&	\\
\hspace{0.1em}\vspace{-0.5em}\\
C4	&	$6.041^{+0.025}_{-0.025}$	&	$2.64^{+3.13}_{-2.64}$	&	&	\\
\hspace{0.1em}\vspace{-0.5em}\\
\midrule
$N_T$ (Total)	&	 \multicolumn{4}{c}{$13.96^{+6.40}_{-5.32}\times 10^{10}$~cm$^{-2}$}\\
\bottomrule
\end{tabular}
\end{table*}

\newpage

\subsection{VCA}

\begin{figure}[h]
    \centering
    \includegraphics[width=\textwidth]{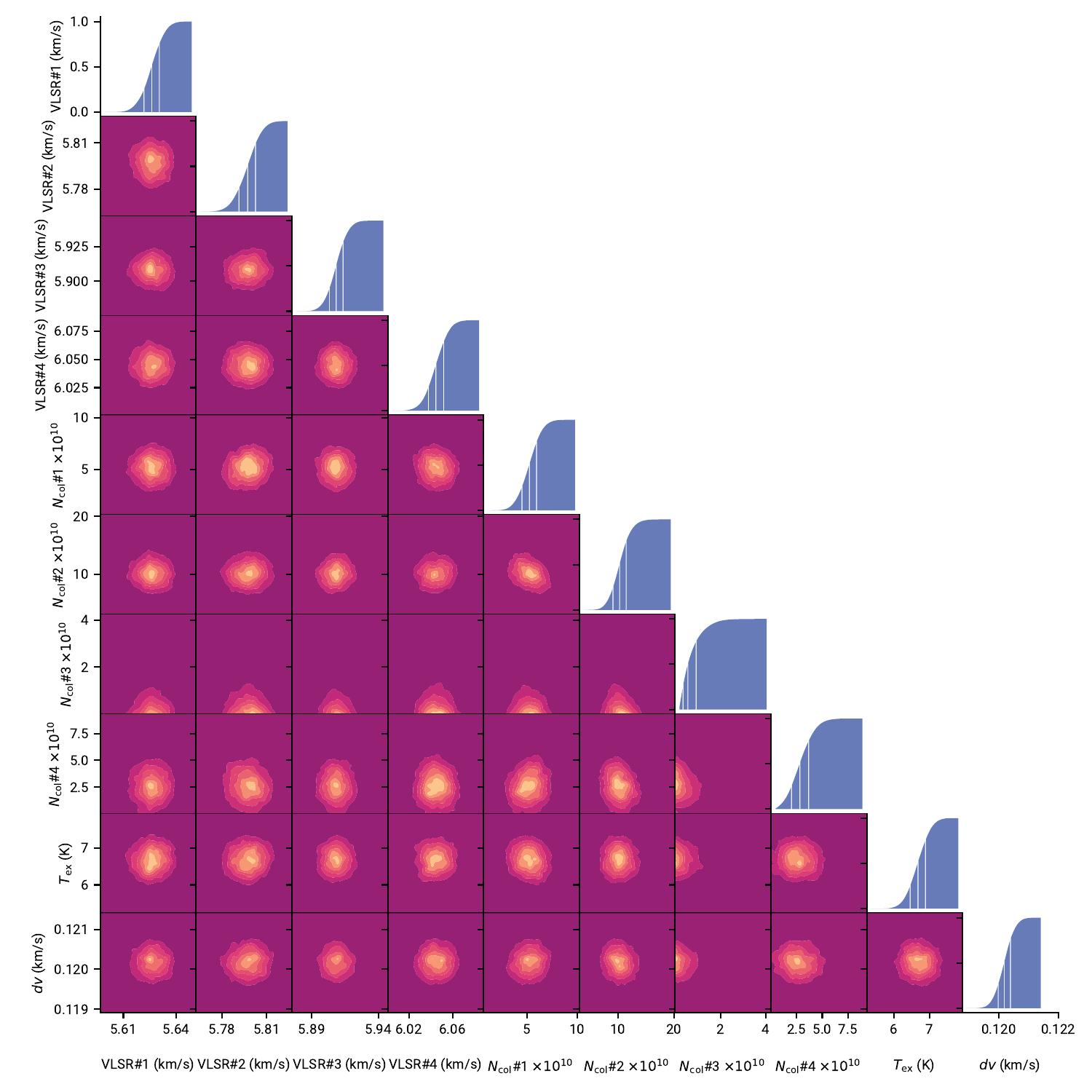}
    \caption{Corner plot for VCA. The diagonal traces correspond to ECDF plots, and off-diagonal plots show the kernel density covariance between model parameters. In the former, lines represent the 25{\nth}, 50{\nth}, and 75{\nth} percentiles respectively. The length scale for the kernel density plots is chosen with Scott's rule.}
    \label{fig:vca-corner}
\end{figure}

Figure \ref{fig:vca-corner} visualizes the MCMC simulation results for VCA, with a similar treatment as to \textit{trans}-(Z)-CVA. In contrast to the other molecules we have studied here, VCA demonstrates significant bimodality in its posterior distributions. Most of the observed flux can be explained by components \#1 and \#4, while our model displays large covariance with component \#3, likely suggesting a three-component model where \#2 and \#3 are degenerate. Summaries of the posterior distributions can be found in Table \ref{table:vca}.

\begin{table*}[hbt!]
\centering
\caption{VCA best-fit parameters from MCMC analysis. Quoted uncertainties correspond to the 95\% highest posterior density. $N_T$ (Total) refers to the pooled column density from the four components, given as the mean and 95\% highest posterior density of the joint posterior distributions.}
\label{table:vca}
\begin{tabular}{c c c c c}
\toprule
\multirow{2}{*}{Component}&	$v_{lsr}$ &	\multicolumn{1}{c}{$N_\mathrm{col}$}					&	$T_{ex}$							&	$\Delta V$		\\
			&	(km s$^{-1}$)				&	\multicolumn{1}{c}{(10$^{10}$ cm$^{-2}$)} 	&	(K) & (km s$^{-1}$)	\\
\midrule
\hspace{0.1em}\vspace{-0.5em}\\
C1	&	$5.626^{+0.013}_{-0.013}$	&	$5.185^{+2.183}_{-2.189}$	&	\multirow{6}{*}{$6.7^{+0.6}_{-0.6}$}	&	\multirow{6}{*}{$0.124^{+0.003}_{-0.003}$}\\
\hspace{0.1em}\vspace{-0.5em}\\
C2	&	$5.797^{+0.017}_{-0.017}$	&	$10.218^{+3.526}_{-3.560}$	&	&	\\
\hspace{0.1em}\vspace{-0.5em}\\
C3	&	$5.908^{+0.016}_{-0.015}$	&	 $0.569^{+1.002}_{-0.569}$	&	&	\\
\hspace{0.1em}\vspace{-0.5em}\\
C4	&	$6.045^{+0.020}_{-0.021}$	&	$2.708^{+2.284}_{-2.474}$	&	&	\\
\hspace{0.1em}\vspace{-0.5em}\\
\midrule
$N_T$ (Total)	&	 \multicolumn{4}{c}{$18.669^{+3.704}_{-3.767}\times 10^{10}$~cm$^{-2}$}\\
\bottomrule
\end{tabular}
\end{table*}

\newpage

\subsection{Ethyl cyanide \label{sec:ethyl}}

As part of our analysis into the hypothesis of cyanopolyyne hydrogenation of \ce{HC5N} leading to the formation of \ce{[H3C5N]} isomers, we performed the same velocity stacking and matched filter analysis for ethyl cyanide, the hydrogenation product of vinyl cyanide, which in turn could be formed from hydrogenation of \ce{HC3N}. The catalog for ethyl cyanide was generated using spectroscopic parameters collated in the Cologne Database for Molecular Spectroscopy \citep{muller_cologne_2005,endres_cologne_2016}.

\begin{figure}[h]
    \centering
    \includegraphics[width=\textwidth]{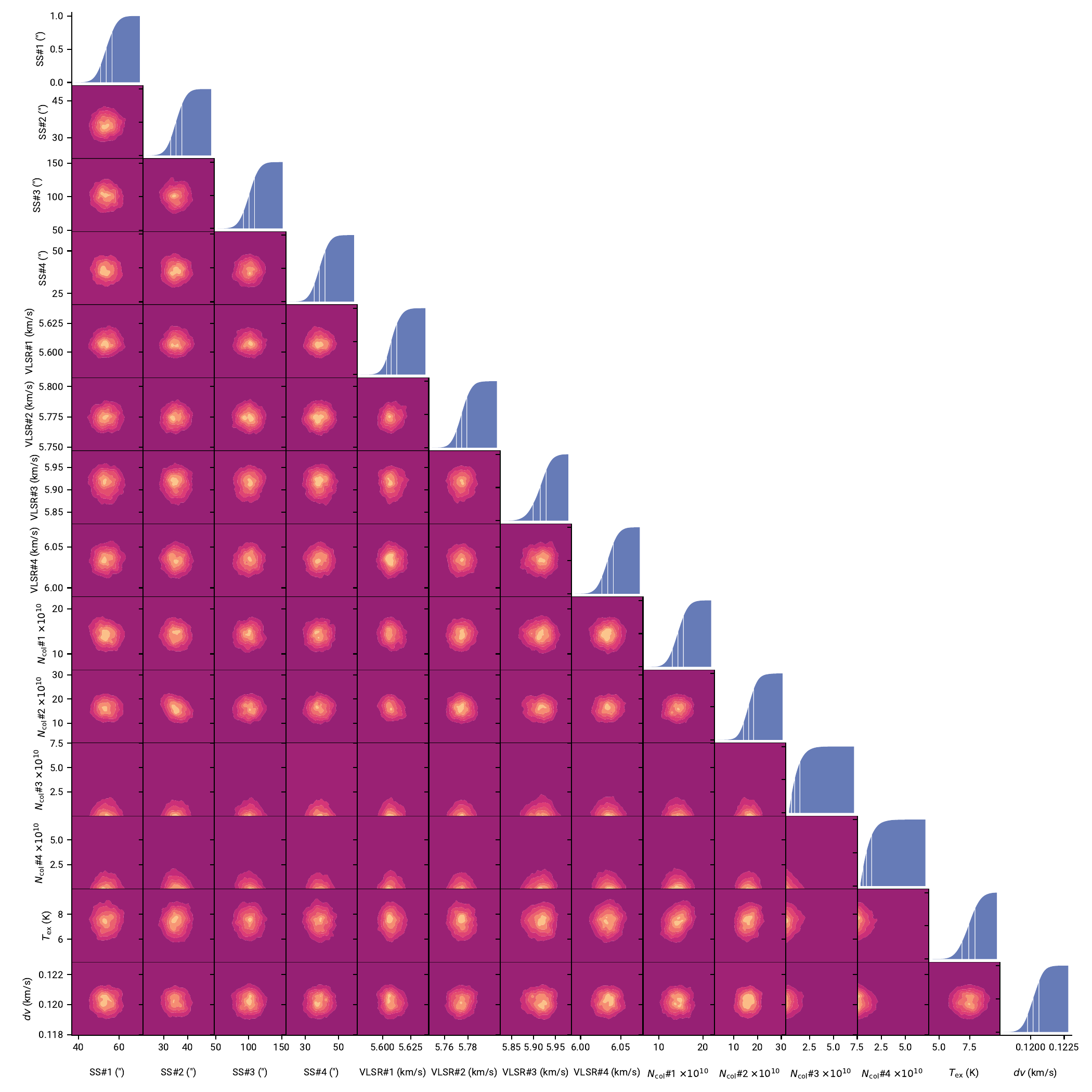}
    \caption{Corner plot for ethyl cyanide. The diagonal traces correspond to ECDF plots, and off-diagonal plots show the kernel density covariance between model parameters. In the former, lines represent the 25{\nth}, 50{\nth}, and 75{\nth} percentiles respectively. The length scale for the kernel density plots is chosen with Scott's rule.}
    \label{fig:ethyl-corner}
\end{figure}

\begin{figure}[h]
    \centering
    \includegraphics[width=0.44\textwidth]{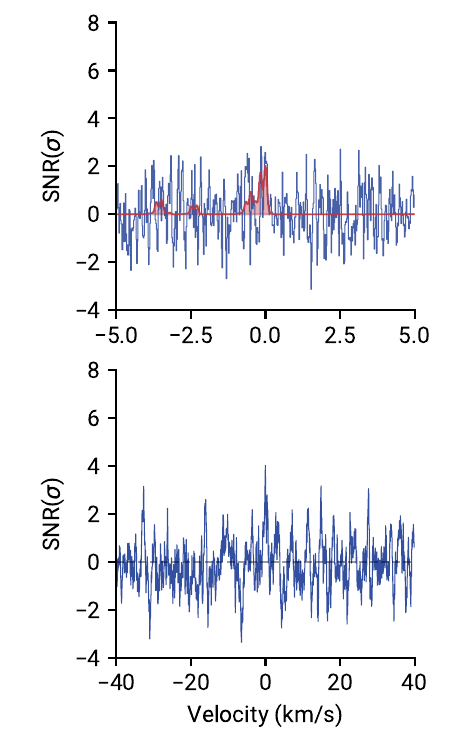}
    \caption{Velocity stack and matched filter plots for ethyl cyanide. The peak matched filter response is 4.17$\sigma$.}
    \label{fig:ethyl-mf}
\end{figure}

Individual transitions of ethyl cyanide were not observed at our current level of integration. Figure \ref{fig:ethyl-corner} shows the corner plot from the MCMC simulations, indicating likely detections toward components \#1 and \#2, and most likely non-detections in \#3 and \#4---a summary of the derived parameters can be found in Table \ref{table:ethyl}. At our level of integration, it appears that there is supporting evidence for a \emph{tentative} detection of ethyl cyanide, albeit at relatively low significance (Figure \ref{fig:ethyl-mf}): as the quality of the GOTHAM spectrum improves, we can revisit this molecule in order to place better constraints on the model parameters, and correspondingly improve the matched filter response. At its current state, we establish an upper limit to the total column density based on the 97.5th percentile value from the joint posterior of $<4\times10^{11}$\,cm$^{-2}$.

\begin{table*}[hbt!]
\centering
\caption{Ethyl cyanide best-fit parameters from MCMC analysis. Quoted uncertainties correspond to the 95\% highest posterior density. $N_T$ (Total) refers to the pooled column density from the four components, given as the mean and 95\% highest posterior density of the joint posterior distributions.}
\label{table:ethyl}
\begin{tabular}{c c c c c c}
\toprule
\multirow{2}{*}{Component}&	$v_{lsr}$					&	Size					&	\multicolumn{1}{c}{$N_\mathrm{col}$}					&	$T_{ex}$							&	$\Delta V$		\\
			&	(km s$^{-1}$)				&	($^{\prime\prime}$)		&	\multicolumn{1}{c}{(10$^{10}$ cm$^{-2}$)}		&	(K)								&	(km s$^{-1}$)	\\
\midrule
\hspace{0.1em}\vspace{-0.5em}\\
C1	&	$5.608^{+0.014}_{-0.013}$	&	$54^{+9}_{-8}$	&	$14.29^{+3.80}_{-3.73}$	&	\multirow{6}{*}{$7.4^{+1.5}_{-1.7}$}	&	\multirow{6}{*}{$0.120^{+0.001}_{-0.001}$}\\
\hspace{0.1em}\vspace{-0.5em}\\
C2	&	$5.760^{+0.018}_{-0.018}$	&	$35^{+7}_{-7}$	&	$16.19^{+6.40}_{-6.33}$	&	&	\\
\hspace{0.1em}\vspace{-0.5em}\\
C3	&	$5.914^{+0.042}_{-0.047}$	&	$100^{+25}_{-25}$	&	$0.86^{+1.38}_{-0.86}$	&	&	\\
\hspace{0.1em}\vspace{-0.5em}\\
C4	&	$6.034^{+0.021}_{-0.021}$	&	$38^{+9}_{-10}$	&	$0.88^{+1.38}_{-0.88}$	&	&	\\
\hspace{0.1em}\vspace{-0.5em}\\
\midrule
$N_T$ (Total)	&	 \multicolumn{5}{c}{$32.05^{+4.07}_{-4.00}\times 10^{10}$~cm$^{-2}$}\\
\bottomrule
\end{tabular}
\end{table*}

\end{document}